\documentclass[10pt]{article}
\usepackage{graphicx}
\usepackage{verbatim}
\usepackage{url}
\begin{document}
\title{Regrets, learning and wisdom}
\author{Damien Challet\\~\\
\small Laboratoire de Math\'ematiques et Informatique \\
\small pour la Complexit\'e et les Syst\`emes\\
\small
CentraleSup\'elec, Universit\'e Paris-Saclay,
France \\ 
\small Encelade Capital SA, Lausanne, Switzerland\\
\small \url{damien.challet@centralesupelec.fr}}
\maketitle
\abstract{This contribution discusses in what respect Econophysics may be able to contribute to the rebuilding of economics theory. It focuses on aggregation, individual vs collective learning and functional wisdom of the crowds. 
} %end of abstract

\section{Introduction}

A good starting point to rebuild an economic theory is to use agent-based models that include learning, interaction and networks \cite{kirman2,bouchaud2008economics,bouchaud2009unfortunate,battiston2016complexity}. This framework is a natural meeting point for Economics and Physics (and Psychology and Biology and Computer Science and ...), which already hints that Econophysics is only part of the solution.
 
Statistical Physics's strength comes from its familiarity with collective phenomena. Aggregating the non-linear actions of many interacting individuals leads to remarkable global phenomena and great mathematical simplifications \cite{MPV,privman2005nonequilibrium}. Whether the outcome is optimal for the agents or the system is of central importance. This contribution argues that learning and optimality may occur at various levels, may be either implicit or explicit, and that Econophysics would be wise to incorporate some more ideas from Neuroscience, Computer Science and Experimental Psychology.

Let me start with a few generic remarks about the difference of approaches to data analysis and modelling in Econophysics and Economics or Finance.

\section{Interdisciplinary communication}

The grass in other fields seems not only greener but also disconcerting at times. This, of course, works both ways. 

\subsection{Statistics}

In their 10-year old worries about Econophysics \cite{gallegati2006worrying}, Gallegati, Keen, Lux and Ormerod pinpointed the general disregard of econophysicists for Statistics,  and rightly so. The spontaneous reaction of physicists is/was to assume to have collected enough data to dispense with statistical tests and tables. The situation has much changed in this respect. Confidence intervals of estimates are not unheard of nowadays. Physicists not only use some statistical tools, but even propose new statistics \cite{tumminello2011identification,malevergne2011testing,challet2015sharperatio,gualdi2016statistically}.

\subsection{Star(t)ling fits}

\begin{figure}
\centerline{\includegraphics[scale=0.5]{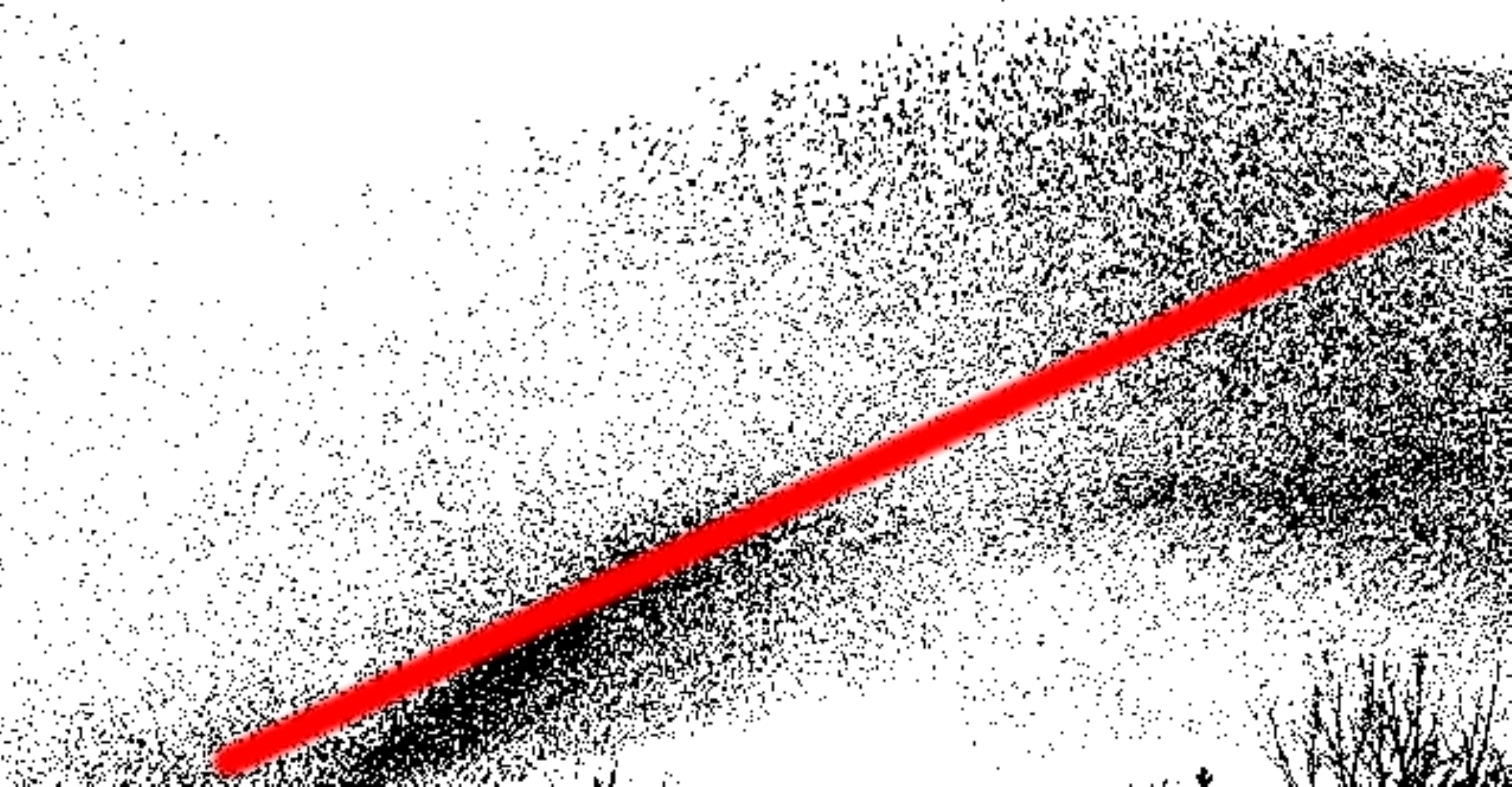}}
\caption{Starlings flocking and a line. Original image source: Wikipedia.org\label{fig_starl}}
\end{figure}

The abundance of linear fits in Economics and Finance papers often puzzles physicists.  Let me discuss what model fitting implies generically. Measuring something is equivalent to projecting a system into a sub-space. A good example is that of a picture taken by a camera: it is projection of a 3+1-dimensional world into a 2-dimensional world. In addition, the position of the camera is also of crucial importance. Figure \ref{fig_starl} shows of a flock of bird. For centuries, people have puzzled about the 3-dimensional structure of such flocks, before it was realized that these clouds were dynamical two-dimensional objects, i.e., ribbons \cite{cavagna2010scale}.  Had anyone been able to ask a bird what the shape was like from the inside, the shape of starlings' flocks would be have been known a long time ago. In short, placing oneself in the right space is a necessary condition for meaningful  fits. 

Taking linear regressions against a few well-known factors is the same thing as taking a picture from a limited number of popular standpoints. Say that one wishes to analyse the performance of a collection of hedge funds. A questionable approach is to project their performance in the space spanned by the few Fama-French factors \cite{famafrench1993} and then argue that hedge funds trade on such and such factors. This is inherently incomplete and very unlikely to yield real understanding of what drives hedge funds performance. What one really needs to do is to reverse-engineer the performance of hedge funds. This requires to place one self in a space that encompasses all likely used trading strategies, or equivalently, to define factors as the returns of a variety of these strategies. Reference  \cite{weber2013hedge} is mostly successful in replicating the returns of several thousand funds, except two: the feeder fund of Madoff, which lived in a fantasy world, and Capital Fund Management Stratus Fund because the chosen strategy space did not include the strategies of that particular fund, which resulted in an meaningless projection.

Fitting any kind of model, however more sophisticated, to data is a  projection. For example, calibrating a simple agent-based model \cite{kirman1991epidemics} to financial data based on a moment method is a double projection, i.e., a double dimensionality reduction: from the data to moments and from the model to moments \cite{alfarano2006estimation}. Model and data meet in a third space. This may yield an incomplete fit, which therefore may not be more efficient than other approaches. This leads us to microscopic models.
 
 %That relationships must be linear, or bilinear must be seen as  local expansions, of limited use.

\section{Lurking Ising models}

Initially, econophysicists tried to apply the models  they were most familiar with to financial or economic situations, which sometimes was hair-raising, even for some physicists. There is indeed no reason why financial markets should be exactly equivalent to a gas of electrons, even if some random power-law exponent seems right. This was deeply worrying. Much has changed since then.

The case of the Ising model is less controversial, if only because it is equivalent to Schelling's model \cite{schelling1971dynamic} and because it is easy to see why it is quite likely to appear in  discrete-choice agent-based models. It also illustrates the variety of what a classical spin may describe in other contexts.\footnote{For a review of the Ising model in Econophysics, see \cite{sornette2014physics} and references therein.}  The simplest idea of course is to map the two possible values of a classical spin to two opposite decisions, which seems natural for investors \cite{CB,iori1999avalanche,bornholdt2001expectation}. Another possibility is to map two alternative possibilities to the two spin values:  Ref.\ \cite{bouchaudopinion} proposes a test for the presence of social imitation in the choice between two alternative possibilities. The key point is that this test is based on exact results from mean-field random-field Ising model and consists in non-linear relationships between two quantities, all of which would be impossible to guess {\em a priori}. In other words, analytically tractable aggregation provides much more than moments of intellectual satisfaction.

Yet, all the above models are built as Ising models from scratch:  actions are directly mapped  to classical spins. Assuming instead that the binary choice is which strategy to use also leads to disordered spin models: agents with very limited possible actions in a complex world are able to optimise a global quantity, which  can be written as a mean-field spin Hamiltonian \cite{MGbook} where the disorder comes from agents' heterogeneity. When binary choices are involved in interacting agent models, it is hard to avoid Ising models.

\section{Learning}

\subsection{Microscopic learning}

Whereas Logit learning is often found in Econophysics literature, surprisingly few other types of learning have been investigated. There lies much potential to establish bridges with other disciplines. First, computer scientists have also applied learning to finance \cite{cover1991universal}. More generically, in a Markovian context, Q-learning rests on the assumption that the system may be classified in a finite number of states by the agents and converges towards the optimal policy \cite{watkins1992q}.  Computer scientists duly applied this scheme to the Minority Game, for example, defining a  state as either which strategy was used, or the previous correct decision. Although this is not discussed in their papers, such dynamics seem to converge to a Nash equilibrium (e.g. \cite{andrecut2001q}). Methods from Statistical Physics are without any doubt able to tackle this kind of learning scheme.

Another central question  is what to learn. The current consensus in Neuroscience is that we learn to regret what we did not do and that a kind of  Q-learning describes well how the brain works \cite{montague2006imaging}. In the context of financial markets, this reinforces bubbles and crashes \cite{lohrenz2007fictive}. Indeed, when the price of an asset has an apparent trend, investors that do own any shares of the said asset regret not to have invested earlier, which triggers their investment.  Reversely, when they are invested in an asset whose price begins to fall, they regret not to have closed their positions earlier.

Finally, Econophysicists have  incorporated remarkably few well-known behavioural biases in their models, as they often assume that agents are risk-neutral. True, there are many reasons why one should use the exponential or logarithmic utility functions with about a ton of salt. However, it is surprising that even Prospect Theory \cite{kahnemanprospect} is nowhere to be seen in our agent-based models. By contrast, in their beautiful paper  \cite{barberis2001prospect}, Barberis {\em et al.} further simplify Prospect Theory with linear approximations, keeping the crucial feature that losses are about twice as painful as gains for agents, and a reference point which is a moving average of past wealth. Prospect Theory hence requires to distinguish between gains and losses in the agent payoff equation updates, which inevitably leads to additional complications. 

Adding a pinch of Prospect theory in our agent-based models is doable. For example, asymmetric gains and losses can be included in the Minority Game and turn out to be another cause for the emergence of large fluctuations  \cite{MGbettin}. There is little doubt that  De Dominicis generating functionals \cite{dedominicis} can accomodate Prospect Theory in more complex agent-based models.

\section{Systemic learning}

The Darwinian force in financial markets that makes them adaptive systems has long been noted \cite{farmerforce,zhangecology,lo2004adaptive}. It does not imply however that the agents themselves are adaptive (e.g. that speculative funds calibrate their strategies in real-time). Indeed, at a global level (and at long time scales), the relative importance of a single agent or of a sub-population of agents may evolve because of competitive processes (for food, wealth, price predictability, etc.): fitness selection is an indirect form of global learning  (see e.g. \cite{galla2009minority}). As a result, even economic systems with zero-intelligence agents undergo a global learning process provided that some kind of selection is performed, as reflected by replicator equations in Evolutionary Game Theory \cite{weibull1995evolutionary}. This does not systematically result in a measurable global optimality.

There are cases, however, where a single global quantity is fairly well optimised, sometimes for obvious reasons (e.g.  financial predictability). This is the Vox Populi \cite{galton1907vox}, or Widsom of Crowds \cite{wisdomcrowds}  effect: the aggregation of inconsistent opinions may lead to consistent aggregate estimates. Although well-known historical examples are really about estimating a single outcome such as the position of a lost craft or the weight of some object, the power of aggregation extends to much more generic situations. 

Ensemble learning such as Random Forests \cite{breiman2001random} apply this idea to classification and regression problems. The latter kind of problem implies that an ensemble of imperfect learners may correctly learn functional relationships, a much more difficult task. The implication for Economics is that the many textbook noiseless ``laws'', for example between price and excess demand, may be valid at an aggregate level. In other words, the clearly too simplistic economic intuition exposed in standard textbooks are in fact quite noisy relationships,. In passing the noise may be due in part to the heterogeneity of economic entities.  A most striking illustration of how to extend these  ``theories'' comes from a work on Marseilles Fish Market \cite{hardle1995nonclassical}: one of its figures plots the price paid for a type of fish as a function of the quantity sold for many transactions. A cloud of point emerges. Only when local averages are taken does emerge a relationship similar to those predicted by usual economic theories. More recently, collective portfolio optimisation in the presence of a complex transaction cost structure was found in brokerage data \cite{Lachapelle2010}. 

The conditions under which collective learning may occur are still unclear and provide a nice challenge for the years to come \cite{lorenz2011social,celis2016sequential} and certainly one which Physicists can contribute to. Finding more examples of wisdom of the crowds will also be part of the fun. Beyond the average behavior, the origin and role of heterogeneity in the dynamics of these systems are complementary research topics.

\section{Conclusion}

The question is not if Economics can become a physical science, but how to make it a science. 20 years of multidisciplinary research have convinced me that it will not be enough to sprinkle economic theory with a few more mechanistic ingredients and to take a slightly less axiomatic approach. A famous speech by Jean-Claude Trichet called for help from a wide range of hard and soft sciences \cite{Trichet2010}. Only the combination of Biology, Experimental Psychology, Computer Science and Physics is likely to make a difference. It is hard to disagree with this point of view  \cite{kirman2,bouchaud2008economics,bouchaud2009unfortunate,sornette2014physics,battiston2016complexity}.

Using common tools and concepts will certainly help achieving better cooperation. Agent-based models, learning and networks certainly qualify as common grounds. A good example of cooperation between economists and physicists is the European CRISIS project, which lead to substantial scientific cross-fertilization.
 More sophisticated tools of Statistical Physics such as generating functionals are nowadays used by mathematical  economists, which is great news, but, expectedly, only by the most mathematically minded, very much as in Physics. Mean-field games \cite{lasry2007mean} will also contribute to establish bridges between Physics and Economics (see e.g. \cite{swiecicki2016schrodinger}). 

 With regards to whether less open-minded economists will accept the resulting new economic thinking, there are many reasons to be optimistic. Each economic crisis is an Economics crisis \cite{kirman2}, and leads to more realistic models. For example, the 2008 crisis has triggered much interest in real networks and self-excited processes (e.g. \cite{lando2010correlation}). 
 
Let us build it and they will use it.

\bibliographystyle{plain}
\bibliography{biblio}
\end{document}